\documentclass[%
 reprint,
 amsmath,amssymb,
 aps
]{revtex4-1}
\usepackage{multirow}
\usepackage{graphicx}% Include figure files
\usepackage{dcolumn}% Align table columns on decimal point
\usepackage{bm}% bold math
\usepackage{amsmath}
\usepackage{amssymb}
\usepackage{mathrsfs}
\usepackage{amsfonts}
\usepackage{color, soul}

\def\oversim#1#2{\lower0.5ex\vbox{\baselineskip0pt\lineskip0pt
                 \lineskiplimit0pt\everycr{}\tabskip0pt
                 \halign{$\mathsurround0pt #1\hfil##\hfil$\crcr #2\crcr\sim\crcr}}}

\begin{document}

\preprint{ARXIV/1310.XXXX}

\title{SUSY explanation of the Fermi Galactic Center Excess \\
and its test at LHC Run-II}

\author{Junjie Cao$^{1,2}$, Liangliang Shang$^{2,3}$, Peiwen Wu$^3$, Jin Min Yang$^3$, Yang Zhang$^3$}

\affiliation{ $^1$ Department of Applied Physics, Xi'an Jiaotong University,
          Xi'an, Shanxi 710049, China \\
   $^2$  Department of Physics,
        Henan Normal University, Xinxiang 453007, China \\
  $^3$ State Key Laboratory of Theoretical Physics,
      Institute of Theoretical Physics, Academia Sinica, Beijing 100190,
      China}

%\date{\today}

\begin{abstract}
We explore the explanation of the Fermi Galactic Center Excess (GCE) in the Next-to-Minimal Supersymmetric Standard Model. We systematically consider various experimental constraints including the Dark Matter (DM) relic density, DM direct detection results and indirect searches from dwarf galaxies. We find that, for DM with mass ranging from $30 {\rm GeV}$ to $40 {\rm GeV}$, the GCE can be explained by the annihilation $\chi \chi \to a^\ast \to b \bar{b}$ only when the CP-odd scalar satisfies $m_a \simeq 2 m_\chi$, and in order to obtain the measured DM relic density, a sizable $Z$-mediated contribution to DM annihilation must intervene in the early universe. As a result, the higgsino mass $\mu$ is upper bounded by about 350 GeV. Detailed Monte Carlo simulations on the $3\ell+ E_T^{miss}$ signal from neutralino/chargino associated production at 14-TeV LHC indicate that the explanation can be mostly (completely) excluded at $95\%$ C.L. with an integrated luminosity of 100(200) fb$^{-1}$. We also discuss the implication of possible large $Z$ coupling to DM for the DM-nucleon spin dependent (SD) scattering cross section, and find that although the current experimental bounds on $\sigma^{\rm SD}_p$ is less stringent than the spin independent (SI) results, the future XENON-1T and LZ data may be capable of testing most parts of the GCE-favored parameter region.
\end{abstract}


\maketitle

\section{\label{intro}Introduction}
As a building block of the universe, Dark Matter (DM) is a focus of current particle physics.
The existence of a Weakly Interacting Massive Particle as a DM candidate
has been indicated by some direct detection experiments like DAMA/LIBRA \cite{DAMA},
CoGeNT \cite{CoGeNT1,CoGeNT2}, CRESST \cite{CRESST} and CDMS \cite{CDMS2}, although these results are not consistent with
each other very well and not supported by other experiments such as Xenon \cite{XENON100} and LUX \cite{lux}.
On the other hand, indirect DM searches also reported some
anomalies. Recent data analyses of the Large Area Telescope (LAT) onboard the Fermi Gamma-ray Space Telescope have shown an excess around $1\sim 4$ GeV in the photon energy spectrum coming from
the Galactic Center \cite{Murgia-Fermi,Goodenough_0910.2998,Hooper_1010.2752,Hooper_1110.0006,
Abazajian_1207.6047,Gordon_1306.5725,Abazajian_1402.4090,Hooper_1302.6589,Hooper_1402.6703,Calore_1409.0042}.
It has been shown that this excess can be well explained by a
$\sim$ 35 GeV DM annihilating $100\%$ into $b\bar{b}$ with a thermal averaged cross section of
about $2\times 10^{-26} \rm{~cm^3/s}$, which is remarkably close to the value
required by the measured relic density $\Omega h^2$ \cite{Hooper_1302.6589,Hooper_1402.6703}.

So far several works have studied such a Galactic Center Excess (GCE) in supersymmetry
\cite{Cerdeno_1404.2572,Berlin_1405.5204,GCE_NMSSM_1406.1181,GCE_NMSSM_1406.6372,GCE_NMSSM_1407.0038,GCE_NMSSM_1409.1573,GCE_NMSSM_1409.7864}.
It was found that after considering the constraints from the  $\Omega h^2$,
the most promising DM annihilation channel is $\chi \chi \to a^\ast \to b \bar{b} $
where $a$ is a CP-odd Higgs boson lighter than about 100 GeV.
In the Minimal Supersymmetric Standard Model (MSSM), due to the mass correlation between the
pseudoscalar and the charged/heavy CP-even scalar, the pseudoscalar is generally heavier than
about 300 GeV and thus cannot explain the GCE.
In the Next-to-MSSM (NMSSM), an extra singlet superfield
$\hat{S}$ is introduced, which results in three CP-even Higgs bosons $h_{1\sim 3}$,
two CP-odd Higgs bosons $a_{1,2}$ and five neutralinos $\tilde{\chi}^0_{1\sim 5}$
(an ascending mass order for the same type of particles is assumed, with $\tilde{\chi}_1^0$
acting as DM and denoted by $\chi$ hereafter) \cite{NMSSM_review_0910.1785}. Since the CP-odd Higgs
boson $a_1$ may be singlet-like and rather light,
light DM pair can annihilate mainly through the $s-$channel mediation of $a_1$ in both the early
universe and today if $m_{a_1}\simeq 2 m_\chi$. As shown in \cite{GCE_NMSSM_1406.6372},
in NMSSM a bino-like DM can explain both the GCE and the $\Omega h^2$ through an off-shell $a_1$, while
a singlino-like DM requires a tuned resonance $2 m_{\chi} \simeq m_{a_1}$ to achieve the same goal.

In this work we intend to interpret the GCE with a $\sim 35 {\rm GeV}$ DM in the NMSSM. Compared with
previous works, we consider more constraints (such as current Higgs data) on the model. We note that taking into account the uncertainties of DM profile in the Galaxy as well as the astrophysical uncertainties of background and foreground will result in a wide range of DM annihilation cross sections which can accommodate the GCE \cite{Hooper_1411.4079,Calore_1411.4647}, e.g.
$\langle \sigma v \rangle |_{v\to 0} = (0.4\sim5.0)\times 10^{-26}\,\rm{cm^3/s}$. However, the observation of dwarf galaxies as well as
other comic ray fluxes such as positron and antiproton are capable of setting upper limits on the cross sections of
DM annihilation into various channels\cite{Ackermann_1310.0828,Geringer-Sameth_1410.2242,Fermi-LAT-2014-Anderson,Fermi-LAT-2014-Matthew-Wood,Bringmann_1406.6027,
Cirelli_1407.2173,Hooper_1410.1527}. Taking DM annihilation into $b\bar{b}$ as an example,
the Fermi observation of dwarf galaxies has required $\langle \sigma v \rangle_{b\bar{b}} |_{v\to 0} \lesssim 1.3
\times 10^{-26}\,\rm{cm^3/s}$ for $m_{\rm DM}=35 \,{\rm GeV}$ \cite{Fermi-LAT-2014-Anderson,Fermi-LAT-2014-Matthew-Wood}.
So if we consider the constraint from the dwarf galaxies and ignore possibly stronger constraints from
other comic ray observations \cite{Bringmann_1406.6027,Cirelli_1407.2173,Hooper_1410.1527},
a reasonable DM annihilation rate dominated by $b\bar{b}$ channel can be chosen to be in the range of
$\langle \sigma v \rangle |_{v\to 0} = (0.4\sim1.3)\times 10^{-26}\,\rm{cm^3/s}$ to explain GCE.
With these considerations we obtained different observations from those in \cite{GCE_NMSSM_1406.6372}, e.g. we found that
a singlino-like DM instead of a bino-like DM is easier to explain both the GCE and the correct $\Omega h^2$.
More importantly, we observed that $\chi \chi \to a_1^\ast \to b \bar{b}$ can not alone explain
both the GCE and the $\Omega h^2$, and in order to get the correct $\Omega h^2$, a sizable
s-channel $Z$ contribution to the early DM annihilation is usually needed. Consequently, the higgsino
mass $\mu$ is upper bounded by about 350 GeV, which will be readily tested at the LHC Run-II through the
trilepton signal of neutralino/chargino associated production. We also discuss another interesting aspect of this
GCE-motivated scenario, i.e. the implication of possible large $Z$ coupling to DM for the DM-nucleon spin dependent
(SD) scattering cross section, which did not receive much attention in previous works.

This paper is organized as follows. In Section II we describe the basic features of NMSSM and our scan strategies. In Section III we present our results and discussions. Finally we conclude in Section IV.

\section{Model and Scan Strategies}

The superpotential of the NMSSM is given by \cite{NMSSM_review_0910.1785}
\begin{eqnarray}
  W^{\rm NMSSM} &=& W_F + \lambda\hat{H_u} \cdot \hat{H_d} \hat{S}
  +\frac{1}{3}\kappa \hat{S^3}, \nonumber
\end{eqnarray}
where $W_F$ is the MSSM superpotential without the $\mu$-term, $\hat{H_u}$ and $\hat{H_d}$ are MSSM Higgs superfields,
$\lambda$ and $\kappa$ are coupling coefficients for Higgs superfields. The corresponding Higgs potential is
then parameterized by soft breaking masses $\tilde{m}_{u,d,s}^2$ for Higgs fields $H_{u,d,s}$
and trilinear soft breaking coefficients $A_\lambda$ and $A_\kappa$. In this framework, the CP-even (odd) Higgs mass
eigenstates are mixtures of the real (imaginary) parts of $H_u$, $H_d$ and $s$,  and the neutralino mass
eigenstates are the mixtures of bino, wino, higgsinos and
singlino. In practice, the parameters $\tilde{m}_{u,d,s}^2$ are traded for $m_Z$, $\tan \beta \equiv
v_u/v_d$ and $\mu \equiv \lambda v_s $ as theoretical inputs.

In our analysis, we fix all soft masses and soft trilinear parameters in squark (slepton) sector at 2 (0.3) TeV except that
we allow the soft trilinear couplings $A_t=A_b$ to vary to obtain a $\sim 125$ GeV CP-even Higgs. In order to get a
light bino-like DM, we abandon the GUT relation among gaugino masses and set wino mass $M_2= 1 ~{\rm TeV}$ and gluino mass $M_3 = 2 ~{\rm TeV}$. Thus the free parameters are $\tan\beta, \mu, \lambda, \kappa, A_\lambda, A_\kappa$ in the Higgs sector, bino mass $M_1$ and $A_t$, which are all defined at 2 TeV. We adopt the Markov Chain Monte Carlo method to scan following parameter space with $\textsf{NMSSMTools-4.3.0}$ \cite{nmssmtools}:
\begin{eqnarray}
&& 1 < \tan\beta < 40,~ 0 < \lambda < 0.7, ~ 0 < |\kappa| < 0.7, \nonumber\\
&& 0 < |A_\kappa| < 2 ~{\rm TeV}, ~ 0 < A_\lambda < 5 ~{\rm TeV}, ~ |A_t| < 5 ~{\rm TeV}, \nonumber\\
&& 0 < |M_1| < 0.6 ~{\rm TeV}, ~ 0.1 ~{\rm TeV}< \mu < 0.6 ~{\rm TeV}.
\end{eqnarray}

We select the samples by  following steps: we require $30 ~{\rm GeV} \leq m_{\chi} \leq 40 ~{\rm GeV}$ and impose
all constraints encoded in the $\textsf{NMSSMTools-4.3.0}$ including the relic density at $3 \sigma$ level
(corresponding to $0.107\leq \Omega h^2 \leq 0.131$), LUX exclusion limit at $90\%$ C.L. and various B-physics
measurements with criteria described in detail in \cite{NMSSM_DM_1311.0678}. Then we use $\textsf{HiggsBounds-4.1.2}$
\cite{HiggsBounds} to systematically impose the constraints from Higgs searches at LEP, Tevatron and LHC. We also
perform a fit to the Higgs data updated in the summer of 2014 with details described in \cite{Cao-New-Higgs-Fit}
and keep the $2\sigma$ samples with $\Delta\chi^2=\chi^2-\chi^2_{min}<6.18$. Subsequently we calculate the DM annihilation
cross section today with $\textsf{micrOMEGAs-3.6.9.2}$ \cite{micromegas} and keep the samples with
$\langle \sigma v \rangle |_{v\to 0} = (0.4\sim1.3)\times 10^{-26}\,\rm{cm^3/s}$ to explain GCE.

%%%%%%%%%%%%%%%%%%%%%%%%%%%%%%%%%%%%%%%%%%%%%%%%%%%%%%%%%%%%%%
\begin{table}[t]
\caption{Expected cross sections of the SM background after cuts for fix signal regions in \cite{atlas_3lepton} and the SRZd region
defined in this work at 14-TeV LHC. }\label{table1}
\begin{tabular}{|c|c|c|c|c|c|c|c|c|}
\hline
\multirow{2}{*}{\begin{tabular}[c]{@{}l@{}} ~ \end{tabular}} & \multicolumn{7}{c|}{Expected cross section (fb)}     \\ \cline{2-8}
                                    & SRnoZa & SRnoZb & SRnoZc & SRZa  & SRZb & SRZc & SRZd \\ \hline
$Z^{(*)}Z^{(*)}$                    & 1.32   & 0.20   & 0.03   & 0.90  & 0.12 & 0.04 & 0.01 \\ \hline
$Z^{(*)}W^{(*)}$         & 4.33   & 1.96   & 0.23   & 22.28 & 2.06 & 0.58 & 0.24 \\ \hline
$t\bar{t}$                          & 4.97   & 1.31   & 0.28   & 0.90  & 0.11 & 0.06 & 0.00 \\ \hline
Total                               & 10.62  & 3.47   & 0.54   & 24.08 & 2.29 & 0.68 & 0.25 \\ \hline
\end{tabular}
\end{table}

We find that the surviving samples are characterized by $ 9< \tan \beta < 33$, $170\, \rm{GeV} < \mu < 350\, \rm{GeV}$
and $m_{h^\pm} > 500 ~{\rm GeV}$. We classify them into four scenarios: for scenario I-S (I-B), $h_1$ corresponds to $h_{\rm SM}$ and DM is singlino(bino)-like, while for scenario II-S (II-B), $h_2$ acts as $h_{\rm SM}$ with DM being singlino(bino)-like.
Since the higgsino mass $\mu$ of the surviving samples is not very large, they should be testable at 14-TeV LHC through the channel $p p \to \tilde{\chi}_i^\pm \tilde{\chi}_j^0 \to 2 \tilde{\chi}_1^0 W Z \to 3\ell+ E_T^{miss}$ where $i=1,2$ and $j=1 \sim 5$\cite{atlas_3lepton}. For each sample we perform a simulation, in which we use MadGraph5 \cite{MadGraph5_2014} and Pythia \cite{Pythia} to generate relevant events and apply the parton shower. With Delphes \cite{Delphes} encoded in CheckMATE-1.16 \cite{CheckMATE}, we obtain the cut efficiencies for the six signal regions (SRs) of \cite{atlas_3lepton}. Then we calculate $\sigma(pp\to\tilde{\chi}_i^\pm \tilde{\chi}_j^0)$ with Prospino2 \cite{prospino} at next-to-leading order and evaluate the significance $\mathcal{S}=s/\sqrt{b+(10\%b)^2}$ for each SR, where $s$ and $b$ correspond to the number of signal and background events after cuts and $10\%$ is the assumed systematical uncertainty of the backgrounds. Moreover, in order to probe the moderately large $\mu$ region more efficiently, besides the six SRs in \cite{atlas_3lepton} we consider one more SR named SRZd, which has the same cuts as SRZc in \cite{atlas_3lepton} except that it requires $E_T^{miss}>165$ GeV. In Table.\ref{table1}, we list our backgrounds after cuts in different SRs. We also performed similar analysis of the trilepton signal at 8-TeV LHC and found no limits on the surviving samples.

\vspace{-0.3cm}
\section{\label{results}Results and Discussion}

{\bf A. Scan results:~~}
Among the four scenarios to explain both the GCE and the $\Omega h^2$
in the allowed parameter space, we find that II-S is most favored, and
II-B and I-B are marginally okay by tuning the relevant parameters (we only get
a few benchmark points after a long time scan), while we cannot find any I-S sample.
We present several benchmark points of the
three available scenarios, and show their details in Table.\ref{table2}.

%%%%%%%%%%%%%%%%%%%%%%%%%%%%%%%%%%%%%%%%%%%%%%%%%%%%%%%%%%%%%%
\begin{table}[t]
\caption{Benchmark points for scenario II-S, II-B and I-B, respectively. Quantities with mass dimension are in unit of GeV and the DM annihilation cross section is in unit of $\rm {cm}^3/s$. The scattering cross sections in direct detection are in unit of pb.}
\begin{tabular}{| c | c | c | c|c|c|c|}
\hline
$m_{\chi}$  & ~$m_{a_1}$~ & ~$m_{h_1}$~ & ~$m_{h_2}$~& ~$m_{h^\pm}$~ & ~$A_t$~ & $M_1$~ \\
\hline
(II-S) 35.1 & 69.1 & 56 & 125 & 4700 & 2360 & -286  \\
(II-B) 40.63 & 81.25 & 56 & 126 & 3500 & 2670 & 43.9 \\
( I-B) 40.88 & 81.74 & 126 & 882 & 4340 & 3690 & 43.4 \\
\hline\hline
$\langle \sigma v \rangle _{b\bar{b}}|_{v\to0}$ & $\lambda$ & $\kappa $ & $\tan \beta$ &  $\mu$ &$A_\lambda$ &$A_\kappa$ \\
\hline
$1.0 \times 10^{-26}$ & 0.43 & 0.034 & 20.3  & 227 & 4830  &-110  \\
$0.6 \times 10^{-26}$ & 0.12 & 0.017 & 12.6  & 256 & 3760 &-61  \\
$0.6 \times 10^{-26}$ & 0.24 & 0.40 & 16.7  &  271 & 3890 & -4.1\\
\hline\hline
$10^{4}\Gamma_{Z,inv}$& $\Omega h^2$ & ${\rm{Br}}_{h_{\rm{SM}}^{inv}}$ &  $10^{4}\sigma^{\rm{SD}}_p $ &$10^{10} \sigma^{\rm{SI}}_p $ &~$|y_{a_1\chi\chi}|$~ &~$|y_{a_1b\bar{b}}|$\\
\hline
 $4$ & 0.13 &  11\%& $3.4$ &$5.0$ &0.048 &0.008\\
$0.1$ & 0.13 &  13\%&  $0.3$ &$3.5$ &0.001 & 0.002\\
$ 0.08$ & 0.13 &  6.0\%& $0.2$ &$1.4$ &0.002 & 0.003\\
\hline
\end{tabular}
\label{table2}
\end{table}

To understand why scenario II-S is most favored by the GCE, we start with an effective Lagrangian \cite{GCE_NMSSM_1406.6372,Izaguirre:2014vva,Arina:2014yna,Boehm:2014hva}
\begin{eqnarray}
-\mathcal{L}_{\rm int}=i y_{a_1 \chi\chi} a_1\bar{\chi}\gamma^5\chi+i y_{a_1 b\bar{b}}a_1 \bar{b}\gamma^5 b,
\label{L}
\end{eqnarray}
where $y_{a_1\chi\chi}$ and $y_{a_1b\bar{b}}$ are Yukawa couplings. The cross section for the annihilation
$\chi \chi \to a_1^\ast \to b \bar{b}$ is given by
\begin{eqnarray}
\langle \sigma v \rangle _{b\bar{b}}|_{v\to 0} \propto \frac{y_{a_1\chi\chi}^2 y_{a_1 b\bar{b}}^2 m_{\chi}^2}{(4m_{\chi}^2-m_{a_1}^2)^2+m_{a_1}^2\Gamma_{a_1}^2}.
\label{sigmaVbb}
\end{eqnarray}
This formula indicates that in order to predict a relatively large $\langle \sigma v \rangle _{b\bar{b}}|_{v\to 0}$, either
$y_{a_1\chi\chi} y_{a_1b\bar{b}}$ takes a sufficiently large value or $m_{a_1}$ approaches to $2 m_\chi$.
In NMSSM, the experimental bounds we considered have limited $a_1$ to be highly singlet-like
(so $\Gamma_{a_1}$ is very small), and $m_{a_2} \gtrsim 500 ~{\rm GeV}$. Consequently,
$y_{a_1\chi\chi}$ in scenario II-S mainly gets contribution from the superpotential term
$\kappa \hat{S}^3$ and is approximately $\sqrt{2} \kappa$ \cite{NMSSM_review_0910.1785}.
In contrast, since DM is bino-like in scenarios II-B and I-B,
$y_{a_1 \chi\chi}$ is suppressed by a factor $\lambda (m_Z \sin \theta_W /\mu)^2 (m_{\chi}/\mu)$
with $\theta_W$ being the weak mixing angle \cite{GCE_NMSSM_1406.6372}.
Since $y_{a_1 \chi\chi}$ in scenario II-S can be much larger than those in scenarios II-B and I-B,
$m_{a_1}$ in scenario II-S may slightly deviate from $2 m_{\chi}$ while being capable of explaining the GCE
so that the theory is less tuned (see Table.\ref{table2}). We emphasize that all the three scenarios require a low $\mu$:
in scenario II-S, a low $\mu$ is needed to predict $m_{h_2} \simeq 125 ~{\rm GeV}$ \cite{Cao-NMSSM},
while in scenarios II-B and I-B, a low $\mu$ is necessary to keep $y_{a_1 \chi\chi}$ moderately large.

Finally we discuss scenario I-S, which is featured by $2 \kappa v_s \sim 35 ~{\rm GeV}$ to get the desired
DM mass and the CP-even singlet Higgs mass $|M_{S,33}| > 125 ~{\rm GeV}$ to ensure $h_1$ being $h_{\rm{SM}}$.
Furthermore, to explain the GCE $\tan \beta \gtrsim 10$ is usually needed to enhance the coupling $y_{a_1 b\bar{b}}$,
and the singlet-like $a_1$ should be around $2 m_{\chi}$, which corresponds to the CP-odd singlet Higgs
mass $|M_{P,22}| \sim 70 ~{\rm GeV}$. However, from the expressions of $M^2_{S,33}$ and $M^2_{P,22}$ \cite{NMSSM_review_0910.1785}
\begin{eqnarray}
M^2_{S,33}=\lambda A_\lambda \frac{v_u v_d}{v_s} + \kappa v_s (A_\kappa+4\kappa v_s), \nonumber \\
M^2_{P,22}=\lambda (A_\lambda+4\kappa v_s) \frac{v_u v_d}{v_s}-3\kappa v_s A_\kappa, \nonumber
\label{massmatrices}
\end{eqnarray}
one learns that the appropriate values of $|M_{S,33}|$ and $|M_{P,22}|$ are very difficult to obtain
simultaneously for a usually negative $\kappa A_\kappa$ (see Eq.(C.10) of \cite{GCE_NMSSM_1406.6372}).
In the rest of this work, we will mainly focus on scenario II-S. For scenario I/II-B, we will only
briefly describe their features since delicate tuning is needed in these scenarios.

%%%%%%%%%%%%%%%% Fig.2 %%%%%%%%%%%%%%%%%%%%%%%%%%%%%%%%%%%%%%%%
\begin{figure}[t]
\includegraphics[width=9cm]{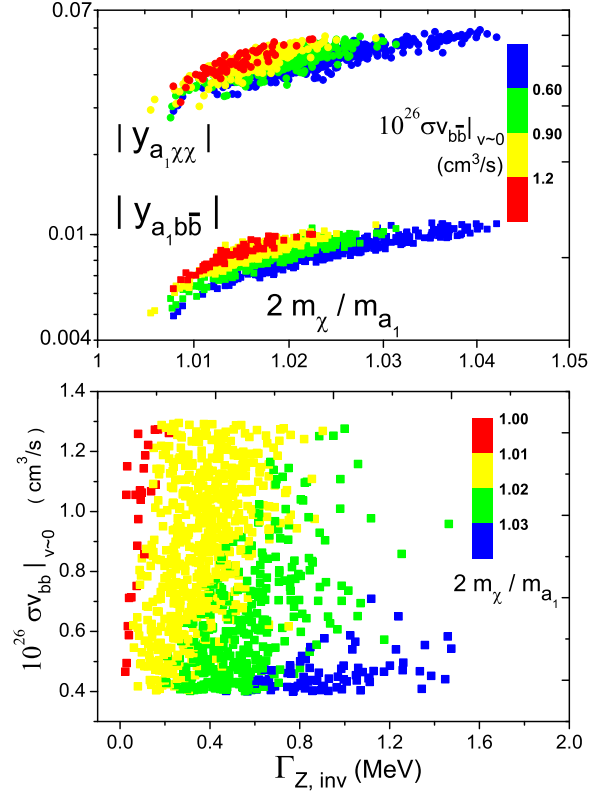}
\vspace*{-1.0cm}
\caption{The dependence of $y_{a_1\chi\chi},y_{a_1b\bar{b}}$ (upper panel), $\langle \sigma v \rangle _{b\bar{b}}|_{v\to0}, \, \Gamma_{Z,inv}$
(lower panel) on $2m_{\chi}/m_{a_1}$ for surviving samples in scenario II-S.}
\label{fig1}
\end{figure}
%%%%%%%%%%%%%%%%%%%%%%%%%%%%%%%%%%%%%%%%%%%%%%%%%%%%%%%%%%%%%%

%%%%%%%%%%%%%%%% Fig.4 %%%%%%%%%%%%%%%%%%%%%%%%%%%%%%%%%%%%%%%%
\begin{figure}[t]
\includegraphics[width=8.7cm]{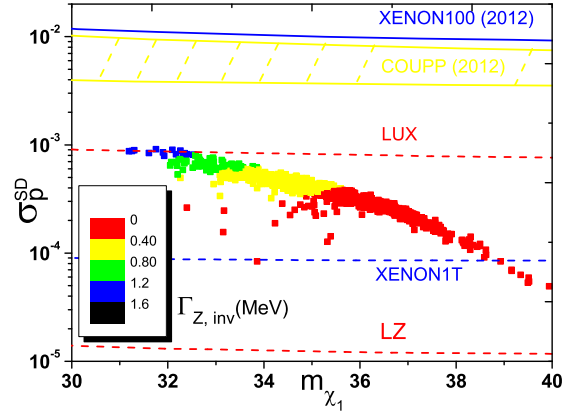}
\vspace*{-1.2cm}
\caption{DM mass $m_{\chi}$ versus DM-nucleon SD scattering cross section $\sigma^{\rm SD}_p$ for II-S samples. Various experimental upper limits are taken from \cite{Cushman_1310.8327}.}
\label{fig2}
\end{figure}
%%%%%%%%%%%%%%%%%%%%%%%%%%%%%%%%%%%%%%%%%%%%%%%%%%%%%%%%%%%%%%

\vspace{0.5cm}
{\bf B. DM annihilation in the early universe:~~}
From Table.\ref{table2} one learns that the three benchmark points all have $2 m_{\chi}/m_{a_1} > 1$, which means that the present DM annihilation $\langle \sigma v \rangle|_{v\to 0}$ is usually larger than that at freezing out $\langle \sigma v \rangle_{0}$ due to the thermal broadening if DM annihilates only through the intermediate $a_1$. However, in order to predict the measured $\Omega h^2$, $\langle \sigma v \rangle_{0}$ should be around the canonical value $3 \times 10^{-26} ~\rm{cm^3/s}$. Since the dwarf galaxy measurements have required $\langle \sigma v \rangle_{b\bar{b}} |_{v\to 0} \lesssim 1.3\times 10^{-26}\,\rm{cm^3/s}$, $\chi \chi \to Z^\ast \to b \bar{b}$ must also contribute sizably to the annihilation in the early universe. A good way of illustrating this is to observe the $Z$ boson invisible decay width into DM pair $\Gamma_{Z,inv}$ which contains the coupling of $Z$ boson to DM pair.

As stated above, in the following we concentrate on the samples of scenario II-S. In the upper and lower panels of Fig.\ref{fig1}, we show $y_{a_1\chi\chi}$ and $y_{a_1b\bar{b}}$ as a function of $2m_{\chi}/m_{a_1}$ and the correlation between $\langle \sigma v \rangle _{b\bar{b}}|_{v\to 0}$ and $\Gamma_{Z,inv}$, respectively. The upper panel indicates that as $2m_{\chi}/m_{a_1}$ increases, $y_{a_1\chi\chi}$ and $y_{a_1b\bar{b}}$ must also increase to maintain an appropriate $\langle \sigma v \rangle _{b\bar{b}}|_{v\to 0}$ to explain the GCE. This panel also shows that for a fixed $y_{a_1\chi\chi}$ or $y_{a_1b\bar{b}}$, generally a $2 m_{\chi}/m_{a_1}$ closer to 1 corresponds to a larger $\langle \sigma v \rangle _{b\bar{b}}|_{v\to 0}$ since the cross section in Eq.(\ref{sigmaVbb}) is very sensitive to the resonance. One can also clearly see that for a fixed $2 m_{\chi}/m_{a_1}$, increasing $y_{a_1\chi\chi}$ and/or $y_{a_1b\bar{b}}$ will help to obtain a larger $\langle \sigma v \rangle _{b\bar{b}}|_{v\to 0}$ as expected.

The lower panel indicates that very close to the resonance region (e.g. red samples with $2 m_{\chi}/m_{a_1}<1.01$) $\Gamma_{Z,inv}$ is quite small, generally in the range of $\Gamma_{Z,inv}<0.2\, {\rm MeV}$, which means $\chi \chi \to Z^\ast \to b \bar{b}$ contribution to $\langle \sigma v \rangle_{0}$ is limited. As $2 m_{\chi}/m_{a_1}$ departs from the resonance one usually has an increased $\Gamma_{Z, inv}$ and decreased $\langle \sigma v \rangle _{b\bar{b}}|_{v\to 0}$, which is most obvious for $2 m_{\chi}/m_{a_1}>1.03$. This is because in the resonance region, the correlation between the $a_1$ contributions to $\langle \sigma v \rangle_{0}$ and to $\langle \sigma v \rangle |_{v\to 0}$ is relatively weak \cite{Abnormal-annihilation} and both can be quite large, in which case the $Z$ contribution to $\langle \sigma v \rangle_{0}$ can be small. When $a_1$ is off-shell, however, the correlation becomes strong and the thermal broadening makes $a_1$ contribution to $\langle \sigma v \rangle_{0}$ get locked to be less than its contribution to $\langle \sigma v \rangle |_{v\to 0}$ which is already smaller than the required canonical value. Consequently, a sizable $Z$ contribution must be present in the early universe. Note that in order to obtain a sizable $Z$ contribution which requires a moderately large coupling $g_{Z\chi\chi}$, $\mu$ can not be too large since $g_{Z\chi\chi} \propto (\lambda v/\mu)^2$ for singlino-like DM and $g_{Z\chi\chi} \propto (m_Z \sin\theta_W/\mu)^2$ for bino-like DM \cite{GCE_NMSSM_1406.6372}. As shown in Fig.\ref{fig1}, a sizable $Z$ contribution allows $2 m_\chi/m_{a_1}$ to deviate moderately from 1 to make scenario II-S less tuned.

Finally we would like to comment briefly on the case of $2 m_{\chi}/m_{a_1} < 1$, where the DM annihilation in the early universe can benefit from the thermal average over the resonance effect of $a_1$. As we mentioned below Eq.(\ref{sigmaVbb}), the highly singlet-like $a_1$ with a very small width will make the resonance effect quite significant \cite{Abnormal-annihilation}. In this case, the relic density forbids the parameter region with very strong resonance, e.g. $0.9 \lesssim 2 m_{\chi}/m_{a_1} \lesssim 1$ \cite{GCE_NMSSM_1406.6372}, and consequently the $a_1$-medicated contribution to $\langle \sigma v \rangle|_{v\to 0}$ is too small to account for the GCE. We checked numerically that,  with the requirement of  a correct relic density for the case of $2 m_{\chi}/m_{a_1}<1$, $\langle \sigma v \rangle|_{v\to 0}$ is generally smaller than $1.0\times 10^{-27}\,\rm{cm^3/s}$.

\vspace{0.5cm}
{\bf C. DM-nucleon scattering:~~}
An interesting implication of sizable $Z$ contribution to $\langle \sigma v \rangle_{0}$ and large $\Gamma_{Z,inv}$ is the DM-nucleon SD scattering cross section $\sigma^{\rm SD}_p$. In Fig.\ref{fig2} we show the DM mass versus $\sigma^{\rm SD}_p$ for II-S samples compared to various experimental upper limits. One can learn that although the current experimental bounds on $\sigma^{\rm SD}_p$ is less stringent than the spin independent (SI) results, the future XENON-1T and LZ data may be capable of testing most parts of the GCE-favored parameter region. One can also notice that an increased singlino-like DM mass $m_{\chi}$ generally correspond to smaller $\Gamma_{Z,inv}$ and $\sigma^{\rm SD}_p$. This is due to the suppressed coupling $g_{Z\chi\chi}\propto (\lambda v/\mu)^2$ as $m_{\chi}\propto 2(\lambda/\mu)^{-1}\kappa$ increases \cite{GCE_NMSSM_1406.6372}, as well as the moderately suppressed two-body decay phase space since $m_{\chi}\sim40\,{\rm GeV}$ is quite close to $m_Z/2$. We confirmed these features numerically for II-S samples.

Note that all of our samples have passed the LUX bounds and since scenario II-S requires $h_2$ to be the SM-like Higgs, one might worry about the possibly large contribution of very light singlet-like CP-even Higgs $h_1$ to DM-nucleon spin independent (SI) scattering cross section. Nevertheless, the singlet-like $h_1$ couples to the singlino-like DM with a coupling $g_{h_1\chi\chi}\sim \sqrt{2}\kappa$ \cite{NMSSM_review_0910.1785} and the light singlino-like DM mass $m_{\chi}\sim 2\kappa (\mu/\lambda)$ requires a small $\kappa$. Consequently, $g_{h_1\chi\chi}$ cannot be very large. We checked that our II-S samples have $\kappa\in(0.02,0.043)$, $m_{h_1}\in (20,100)\,{\rm GeV}$, and for very light $h_1\lesssim 30\,{\rm GeV}$ (with only a small number) $\kappa$ decreases rapidly. Consequently, it is not difficult for scenario II-S to satisfy the direct detection experiment.

%%%%%%%%%%%%%%%% Fig.3 %%%%%%%%%%%%%%%%%%%%%%%%%%%%%%%%%%%%%%%%
\begin{figure}[t]
\includegraphics[width=9cm]{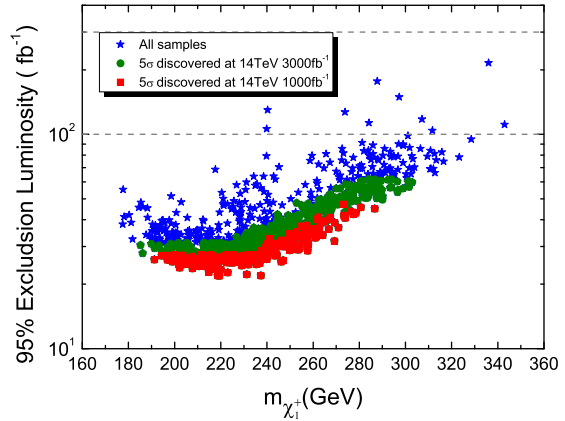}
\vspace*{-0.7cm}
\caption{Required luminosity to exclude the samples of scenario II-S at $95\%$ C.L. at 14-TeV LHC. Samples marked with red square and dark
green bullet may be discovered with an luminosity of 1000 fb$^{-1}$ and 3000 fb$^{-1}$, respectively.  }
\label{fig3}
\end{figure}
%%%%%%%%%%%%%%%%%%%%%%%%%%%%%%%%%%%%%%%%%%%%%%%%%%%%%%%%%%%%%%

\vspace{0.5cm}
{\bf D. Test the explanation at 14-TeV LHC:~~}
Now we discuss the capability of 14-TeV LHC to test the GCE explanation in NMSSM. In Fig.\ref{fig3} we show the needed luminosity to exclude the II-S samples at $95\%$ C.L. as a function of the lightest chargino mass $m_{\tilde{\chi}_1^\pm}$. For each sample, we choose its most sensitive SR, which is usually SRZc for $m_{\tilde{\chi}_1^\pm} \leq 230 ~{\rm GeV}$ and SRZd for $m_{\tilde{\chi}_1^\pm}
\geq 280 ~{\rm GeV}$, and require the corresponding $\mathcal{S}$ (see Sec.II) to be 1.96 to get the exclusion luminosity. Fig.\ref{fig3} indicates that with an integrated luminosity of 100 (200) fb$^{-1}$, most (all) of the II-S surviving samples will be excluded. In Fig.\ref{fig3} we also use red squares and dark green bullets to indicate samples that may be discovered at 14-TeV LHC with 1000 fb$^{-1}$ and 3000 fb$^{-1}$ luminosities, respectively.
As for the benchmark point of scenario II-B (I-B), we find that the exclusion luminosity is 35.2 fb$^{-1}$ (23.7 fb$^{-1}$) and the discovery luminosity is 950 fb$^{-1}$ (300 fb$^{-1}$). Note that, if the trilepton signal is combined with the 2-lepton+jets signal of the $\tilde{\chi}_i^\pm \tilde{\chi}_j^0$ associated production processes as done in \cite{ATLAS-2-Lepton}, the needed luminosity may be further reduced.

\vspace{-0.3cm}
\section{Conclusion}
We scanned the NMSSM parameter space by considering various experimental constraints to explain both the GCE and the measured $\Omega h^2$ with a DM satisfying
$30 ~{\rm GeV} \leq m_{\chi} \leq 40 ~{\rm GeV}$. We have the following observations: a) The GCE can be explained by
the DM annihilation $\chi \chi \to a_1^\ast \to b \bar{b}$ near the resonance region $2 m_\chi/m_{a_1}\sim 1$, and a singlino-like DM is more favored than a bino-like DM; b) When $2 m_\chi/m_{a_1}$ moderately deviates from the resonance, in order to produce the measured relic density, a sizable $Z$ boson contribution to the DM annihilation in the early universe must be present, resulting in the higgsino mass $\mu$ upper bounded by about 350 GeV; c) Although the current experimental bounds on DM-nucleon spin dependent scattering cross section $\sigma^{\rm SD}_p$ is less stringent than the spin independent results, the future XENON-1T and LZ data may be capable of testing most parts of the GCE-favored parameter region; d) Detailed simulations on the $3\ell+ E_T^{miss}$ signal from neutralino/chargino associated production at 14-TeV LHC indicate that the surviving samples can be mostly (completely) excluded at $95\%$ C.L. with an integrated luminosity of 100 (200) fb$^{-1}$, or a large portion of them may be discovered with an integrated luminosity of 3000 fb$^{-1}$.

Finally, we have two comments about our discussions. One is that if we slightly relax the constraint from dwarf galaxies, e.g. to be $\langle \sigma v \rangle |_{v\to 0} \lesssim 2.0 \times 10^{-26}\,\rm{cm^3/s}$, we find that our conclusions keep unchanged. The other one is that when we finished this work, a sophisticated analysis of the
Fermi-LAT data was performed including an estimate of systematic uncertainties \cite{Calore_1409.0042,Calore_1411.4647}. In such a case, the favored DM mass range becomes heavier and wider than previous discussions. We will keep a close eye on the progress in this direction.\newline

This work was supported by the ARC Center of Excellence for Particle Physics at the Tera-scale, by the National Natural Science Foundation of China (NNSFC) under grant No. 10821504, 11222548, 11305049 and 11135003, and also by Program for New Century Excellent Talents in University.

\end{document}